# Klangfarbenakkord and Klangfarbenharmonien

Metric Space Models for Music on Informational Geometry 1

Yusei TAMURA (1, Shigekazu ISHIHARA (2 and Ken ITO (1

(1 The University of Tokyo

(2 Hiroshima International University

Abstract

This paper deals with the introduction of 'geometric harmony', a discipline that explicitly addresses the spectral characteristics of musical gamut. The framework of Western music, from Renaissance to the present, represents sound in terms of 'pitch'—as is evident from its five-line staff notation system—and employs the fundamental frequency as its representative value, 440 Hz, etc. In this paper, by taking the timbres of specific individual instruments as elements and examining the WASSERSTEIN DISTANCE between two voices, and WASSERSTEIN DEVIATIONS between three or more voices, we demonstrate that it is possible to expand the system whilst retaining the entire framework of conventional music theory. At the same time, as an example of practical utility in ensemble playing, we provide a detailed account of the two-voices AFFINITY of the 'Throat G' on clarinet, a note known for its fragility in ensemble contexts.

***Keywords:*** *Chord, Harmony, Spectrum, Probability Density, Wasserstein Metric, Persistent Homology, Stochastic Variance*

## General Overview

In his book "Harmonielehre" (1911/22) [1], Arnold SCHÖNBERG refers to the concept of 'Klangfarbenmelodie'; melody of timbres.

*... I cannot necessarily accept the distinction between timbre and pitch, as it is usually put. I believe that a sound makes itself felt through its timbre, of which pitch is one dimension. Timbre is therefore the broader realm; pitch is one aspect of it.* (...)

*A melody of timbres! What refined senses are able to distinguish this, what a highly developed mind is able to take pleasure in such subtle things!*

*Who would dare to demand a theory here?*

=========================================================

1) 7-3-1 Hongo, Bunkyo Ward, Tokyo, JAPAN

2) 555-36 Kurose Gakuen-dai, Higashi-Hiroshima, Hiroshima JAPAN

Contact address: itosec@iii.u-tokyo.ac.jp

Schönberg himself put forward his well-known ‘dodecaphonic technique’ with the intention of answering such questions. However, the ‘dodecaphony’ in question is nothing more than a sequence of ‘pitch’, not timbre, and do not actually provide an answer to the question posed above.

After the Second World War, this question was broadened, and systematic control of new parameters such as ‘timbre’, ‘speech’ and even ‘spatiality’ was explored (Total Serialism / Schönberg’s Three Questions).

However, during the period from the 1950s to the 1970s, it proved difficult to elucidate these issues due to the analogue technical limitations of sound processing. Furthermore, from the 1980s onwards, as digitalization advanced, interest in these fundamental questions waned; instead, issues directly linked to “musical” “performance” practice—such as the utilization of IRCAM systems—were prioritized, and the fundamental questions were effectively set aside [2].

The aim of this paper is to further expand upon Schönberg’s question theoretically from the standpoints of composers and players, provide an answer based on the level of human understanding achieved in the 2020s, and present concrete examples that are relevant to music.

## 1 Stochastic Spectra and transport problem

In real-world everyday environments, the sounds that reach our ears contain a variety of components. Whilst these are physical sound waves that fluctuate in complex ways over time, the auditory organs of living organisms do not possess physiological functions such as the Fourier transform.

Living organisms, including humans, perform frequency resolution of traveling waves while they are still propagating within the cochlea. This dynamic process separates stimulus components before they are transmitted to the cochlear nerve.

It is difficult to deal with this directly. So, we shall replace it with a short-time Fourier transform below and consider the spectrum of the input stimulus sound.

As Schönberg also pointed out, the physical sound waves reaching the ear are complex and vary over time, and the ‘pitch’ is merely one very limited attribute. A typical example of this is the ‘Edge’ perception of band noise [2], discovered by von Bekesy in 1962–63.

CHEN and ITO have examined the Edge listening in detail, with regard to bandpass-filtered band noise [3]. Pitches within a frequency band slightly inside the passband edges are perceived, whilst the region in between the Edges remains unperceived.

As the perceived pitch is a subjective quantity and follows a Poisson distribution, it is appropriate to regard these as stochastic events (Stochastic spectra) [4].

We shall therefore consider the short-time Fourier transform of a given instrumental timbre or speech signal, and regard the normalized spectrum as the probability density function of the listener’s perception of that frequency when hearing it, and proceed with the following discussion.

Fig. 1-a shows the sonagram of a vocal performance modelled on the ‘yodel’, known as a folk song from the Alpine regions of Switzerland and Austria.

Yodeling is characterized by a shift to falsetto, and this vocal performance exemplifies that style. In contrast, Fig. 1-b shows an example of singing performed at almost the same pitch but by chest voice, without this ‘shift to falsetto’.

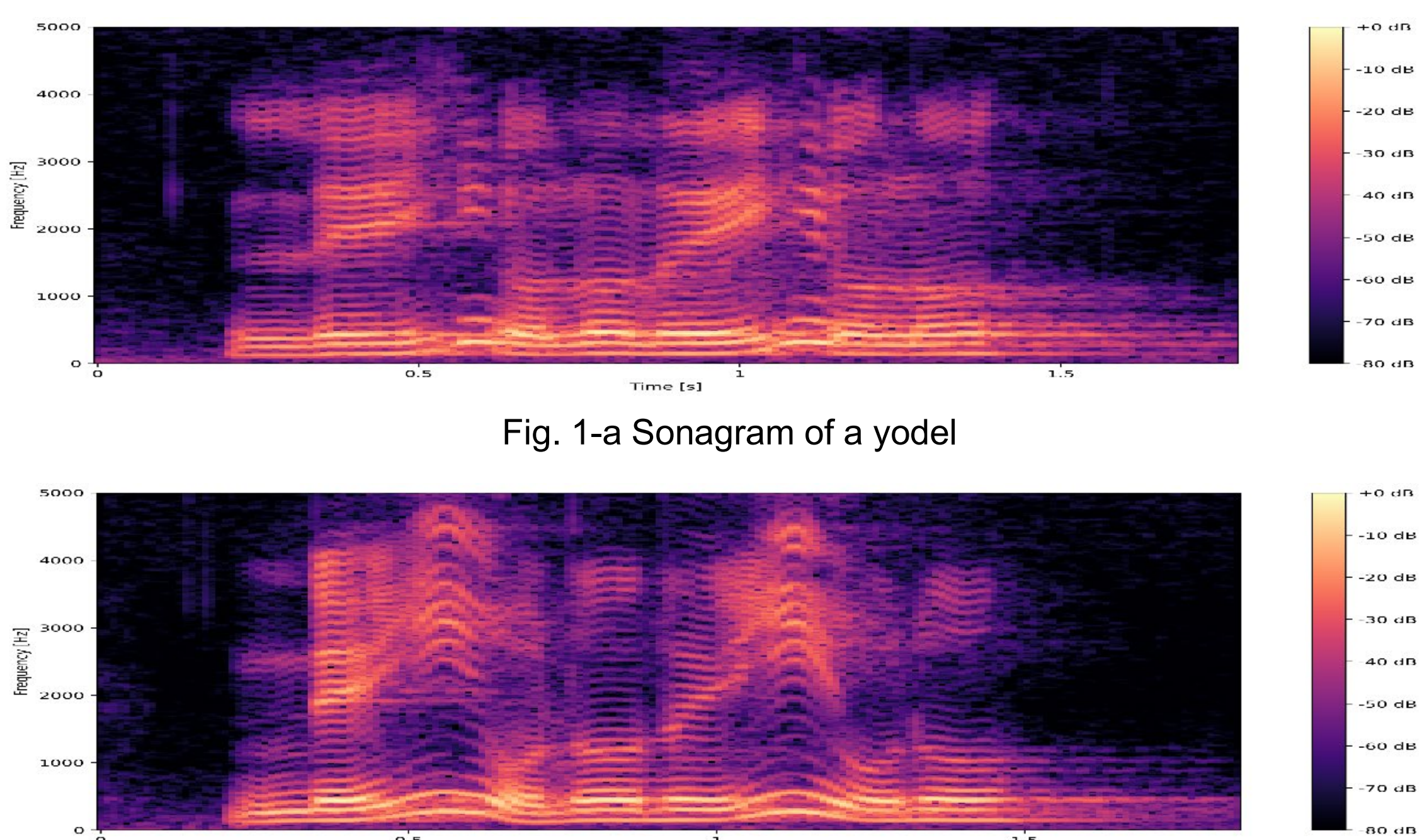

Fig. 1-a Sonagram of a yodel

Fig. 1-b Sonagram of the same pitch sung in chest voice, almost identical to Fig. 1-a

The spectra for each are shown in Figs. 2-a and 2-b.

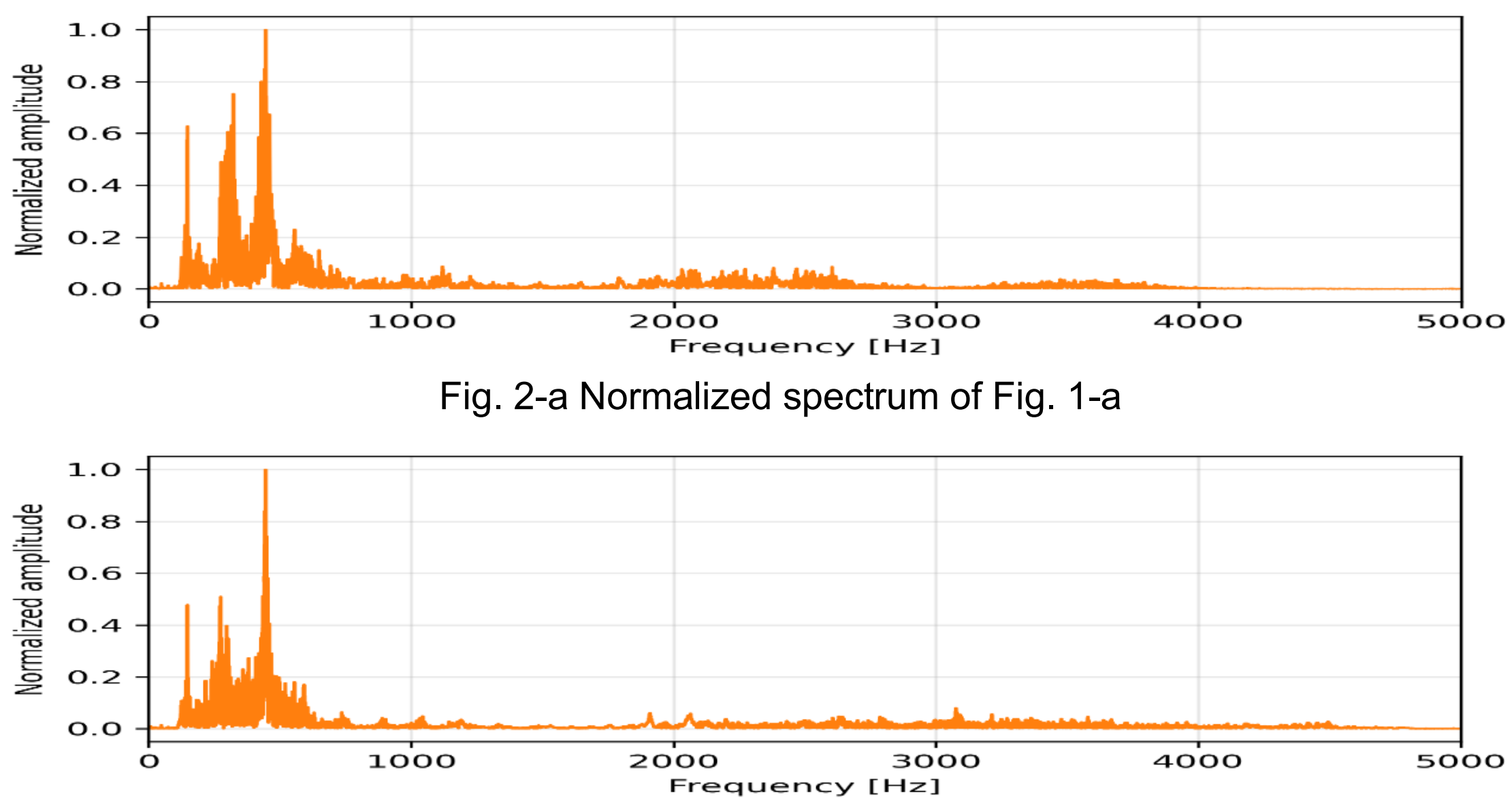

Fig. 2-a Normalized spectrum of Fig. 1-a

Fig. 2-b Normalized spectrum of Fig. 1-b

We now consider transferring the spectrum of the 'chest voice' shown in Fig. 2-b to the 'yodel(falsetto)' in Fig. 2-a, following the optimal transport procedure outlined below. Let us denote the spectra by $s_i$ and $s_j$ and denote the frequency arguments of the element spectra obtained by the Fast Fourier Transform as $f_k$, $f_l$ etc., and define the one-dimensional Wasserstein distance as follows.

$$W_1(s_i, s_j) = \int \left| \sum_{k=1}^{x} s_i(f_k) - \sum_{k=1}^{x} s_j(f_k) \right| \mathrm{d}x$$

・・・①

As this is a one-dimensional transport function in frequency space, the dimension of the Wasserstein distance is [Hz]. Considering the yodel example mentioned earlier, specifically,

$$\begin{aligned} W_1(s_{\text{chest}}, s_{\text{falsetto}}) &= \int_{\text{data length}} \left| \sum_{k=1}^{x} s_{\text{chest},k} - \sum_{k=1}^{x} s_{\text{falsetto},k} \right| \mathrm{d}x \\ &\simeq \int_{\text{data length}} \left| \left( \int_0^x s_{\text{chest}}(k)\, \mathrm{d}k \right) - \left( \int_0^x s_{\text{falsetto}}(k)\, \mathrm{d}k \right) \right| \mathrm{d}x \\ &\approx 442 \text{ [Hz]} . \end{aligned}$$

This results in the following. A separate calculation yields that the spectral centers of mass for each are

$$G_{\text{chest}} \approx 2066, \qquad G_{\text{falsetto}} \approx 1713 \text{ [Hz]} .$$

Therefore, the spectral center-of-mass-shift is estimated to be

$$\Delta G_{\text{chest} \sim \text{falsetto}} \approx 353 \text{ [Hz]} .$$

The Wasserstein distance is always equal to or greater than the spectral center-of-mass shift [5], and since this increment is obtained by altering the 'shape' of the spectrum, the net change in timbre is given by

$$W_1(s_{\text{chest}}, s_{\text{falsetto}}) - \Delta G_{\text{chest} \sim \text{falsetto}} \sim 89 \text{ [Hz]} ,$$

yielding a scalar value in Hz.

It is extremely convenient to be able to estimate changes in timbre in this way, in terms of frequency , and, as we shall see below, it is highly useful as it represents something that has never before existed from the perspective of practical music and musicians [6].

## 2. The Wasserstein distance matrix and the degree of affinity between two spectra

Using the Wasserstein distance, we consider recording the sounds ranging from the lowest to the highest register of a single instrument.

Musically speaking, there are numerous cases—such as ‘alternate fingerings’, ‘harmonics’ and ‘overtone series’—where notes written on a score may have the same pitch but different spectra; we shall discuss typical examples of these in the next section.

Fig. 3-a shows the ‘Wasserstein self-distance matrix’ for the flute, and Fig. 3-b for the viola; these are matrices summarizing the mutual Wasserstein distances calculated for spectral sequences of sample sounds played using standard fingering. Dark blue indicates a low Wasserstein distance value, whilst green and yellow indicate high values; in other words, this suggests low spectral similarity and a tendency for the spectra to diverge. In the case of the flute, a distinct change (register shift) can be observed where the harmonic series changes; however, no such structure is observed in the viola, where finer changes in spectral similarity, such as those caused by changing strings, can be observed.

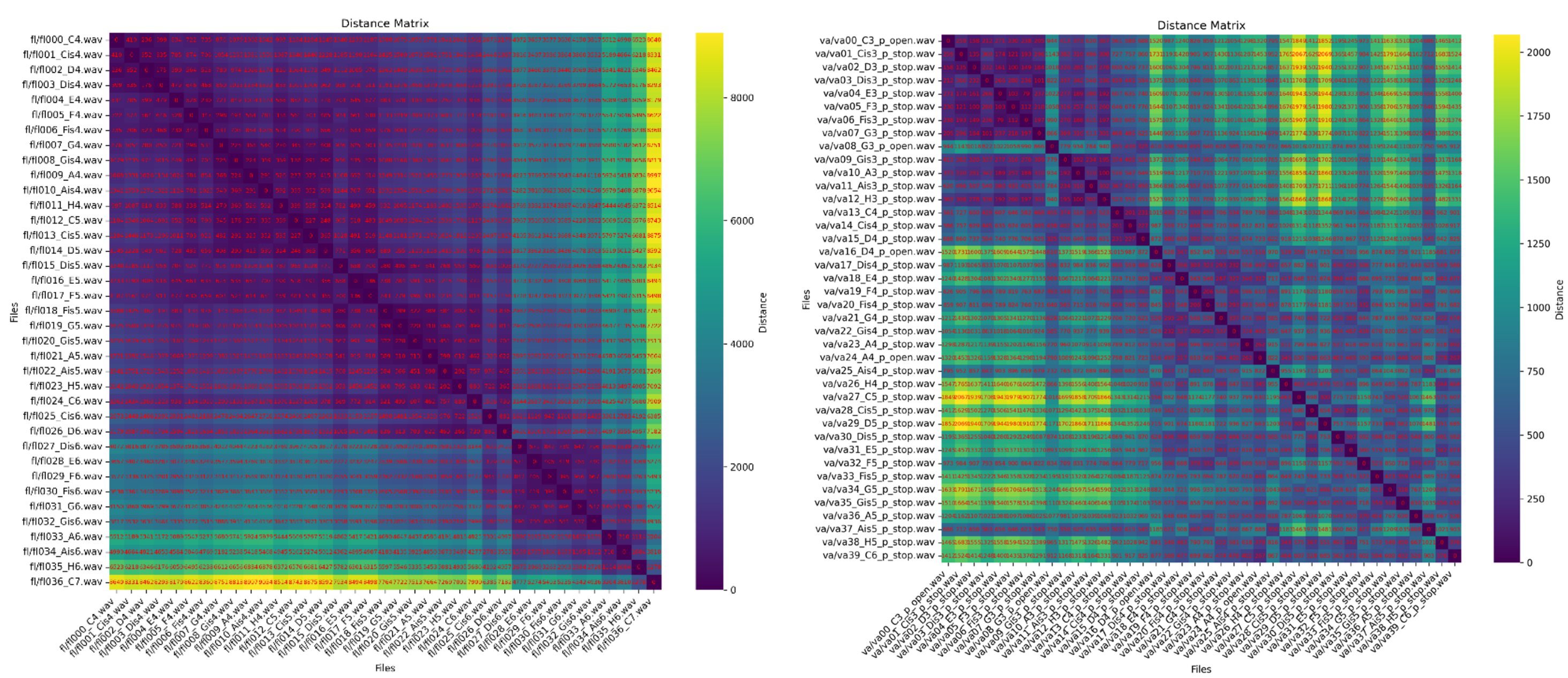

Fig. 3-a: Self-distance matrix for the flute

Fig. 3-b: Self-distance matrix for the viola

We can also take the direct product of the two. By considering the Wasserstein distance matrix for the vectors representing the direct products of the timbres of the flute and the viola, we can define the ‘extended mutual distance matrix for the full-range spectra of the flute and the viola’.

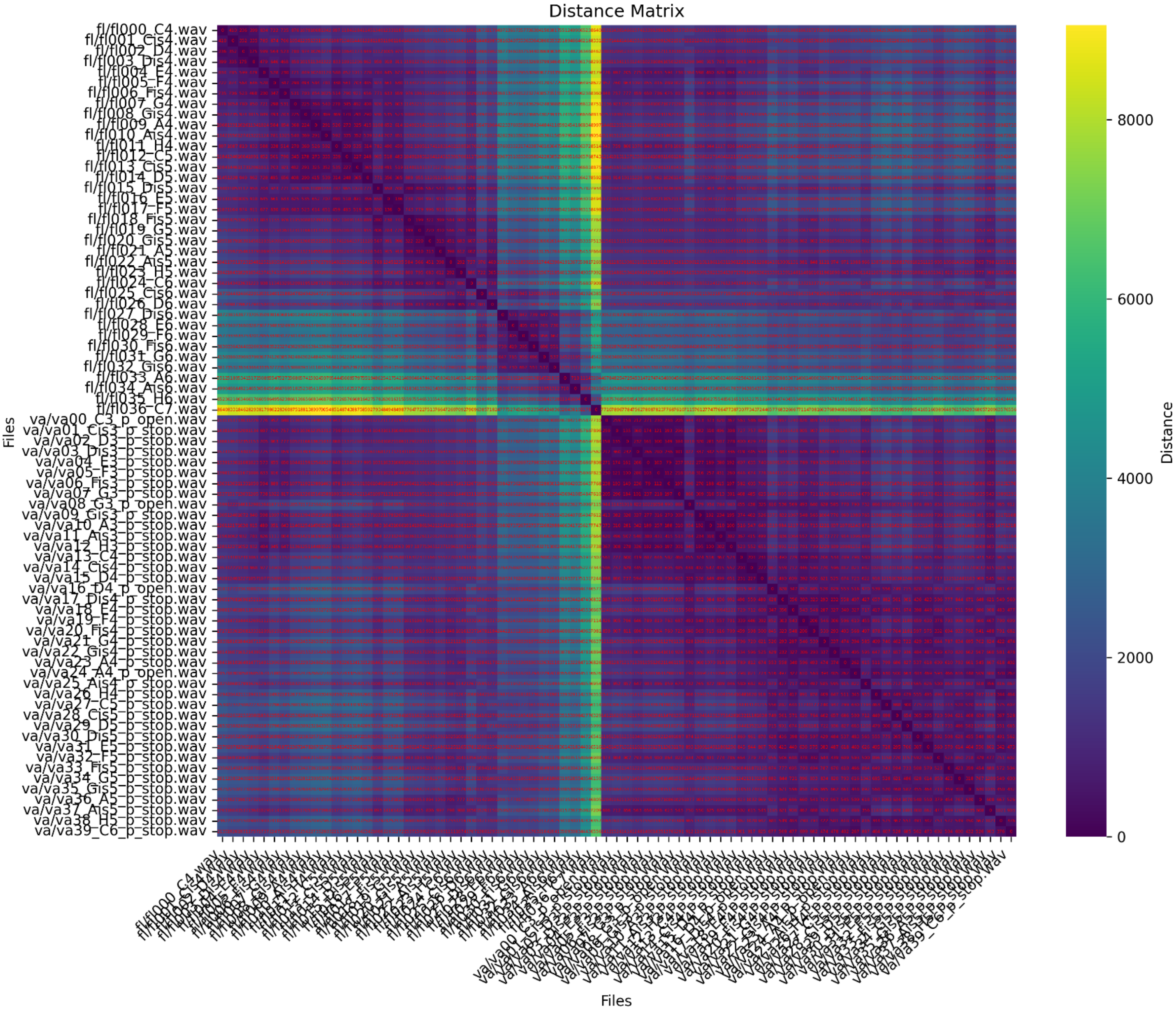

Fig. 4: Expanded mutual distance matrix for the flute and viola

When viewed at a macro level, the diagonal blocks of the expanded matrix contain the self-distance matrix elements, whilst the off-diagonal blocks contain the mutual distance matrix elements. Whilst it is not necessary to construct an expanded matrix if one is simply examining mutual Wasserstein distances, arranging the data in this way results in a symmetric matrix. Although not covered in this paper, this enables a "principal component analysis-style" approach involving diagonalization and the use of eigenvalues and eigenvectors [7].

Returning to the topic of the self-distance matrix for a single instrument, let us examine the timbre of the clarinet.

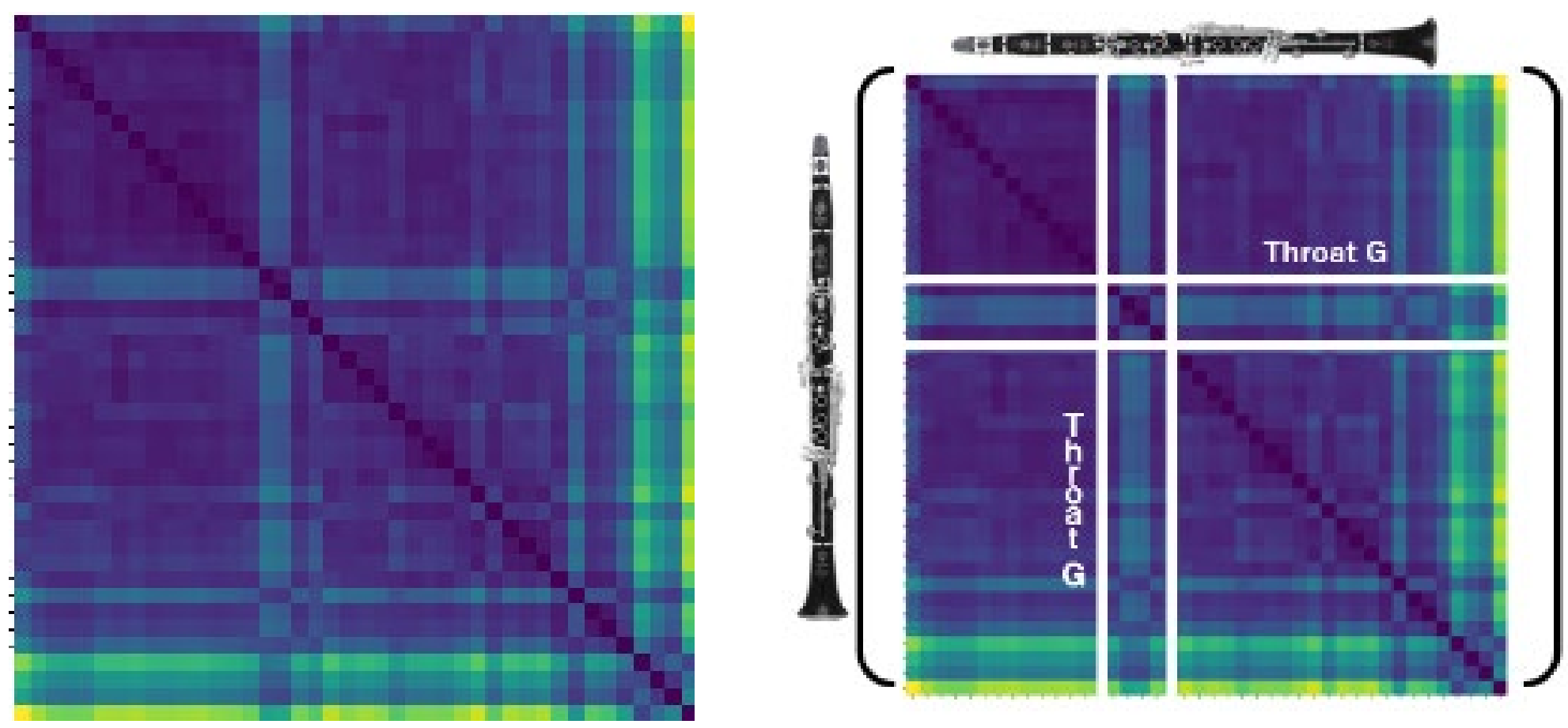

Fig. 5: Self-Wasserstein distance matrix and Throat G for the clarinet

Whilst many notes in the lower register appear deep blue, some areas of lighter color can be observed. This indicates a structural challenge inherent to the clarinet, known as the 'Throat G'. In this register, almost all finger holes are left open. Consequently, the instrument is difficult to control and blends poorly with other instruments, making it challenging to achieve a sense of harmonic unity [Fig. 6].

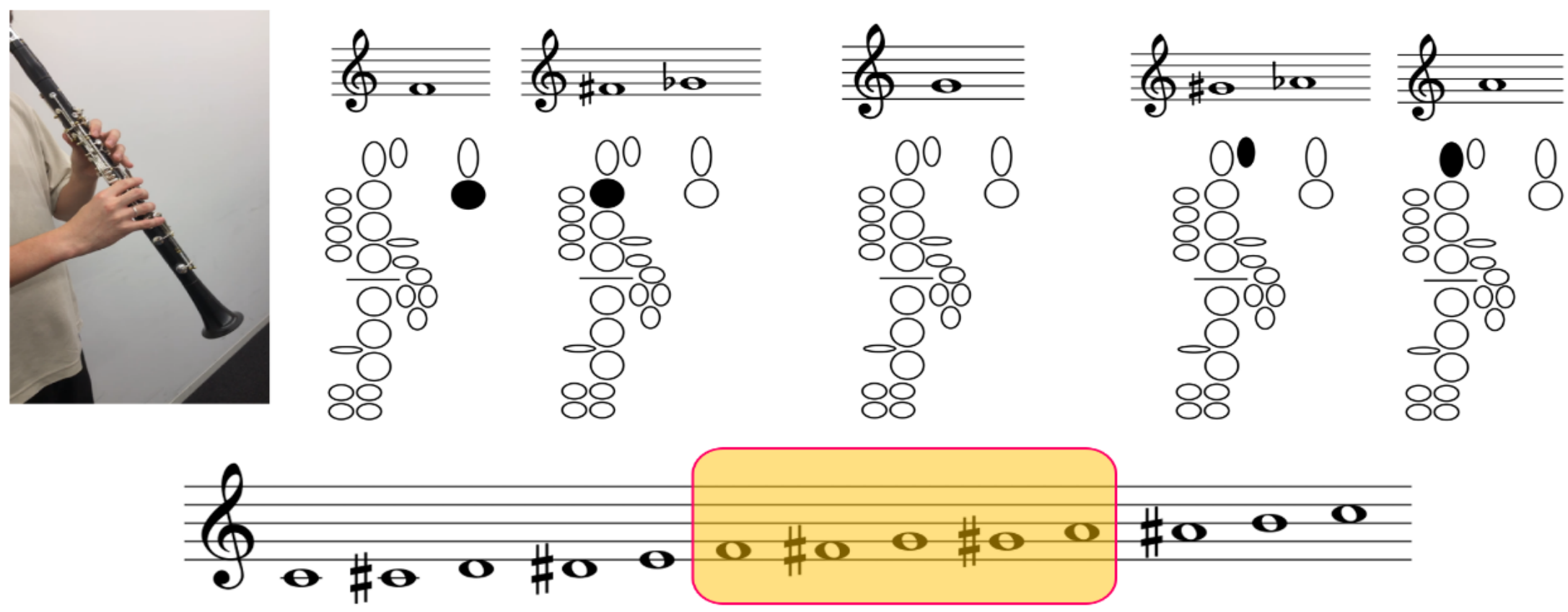

Fig. 6 Fingering around the throat G of the clarinet.

It is not only the note 'G'—played with all finger holes fully open—but also the spectra in the surrounding scale range, from 'F' to 'A' or thereabouts, that stands out clearly from the sounds in other registers. Similar characteristics can be observed in various instances with instruments such as flute's, oboe's and bassoon's family.

The fact that the surrounding timbres have a large Wasserstein distance and are spectrally difficult to blend is not necessarily a musical disadvantage. It should be noted that, given that these notes are physically soft yet tend to stand out clearly, this provides information that is of daily practical use in performance contexts—such as the adoption of fingering patterns that create spectral divergence to ensure that solo instruments or soloistic phrases are not masked by the accompaniment.

## 3. Topological mapping and examples of its application in ensemble

Let us now consider a geometric representation of the Wasserstein distance, which has just been displayed as a distance matrix. In what follows, we will examine topological mapping using Persistent Homology, a technique utilized in fields such as materials engineering and polymer science.

Topological mapping using Persistent Homology has become an established method for extracting and visualizing 'connections between points' from multidimensional data across different scales.

Below, we attempt to visualize the self-distance matrix shown in Fig. 5 for a B♭ or A clarinet whilst preserving the Wasserstein distance. When the spectrum ranging from the lowest to the highest register is arranged using Persistent Homology topological mapping whilst preserving the Wasserstein distance, timbres that are close to one another form clusters, allowing the timbres to be classified by register (Fig. 7).

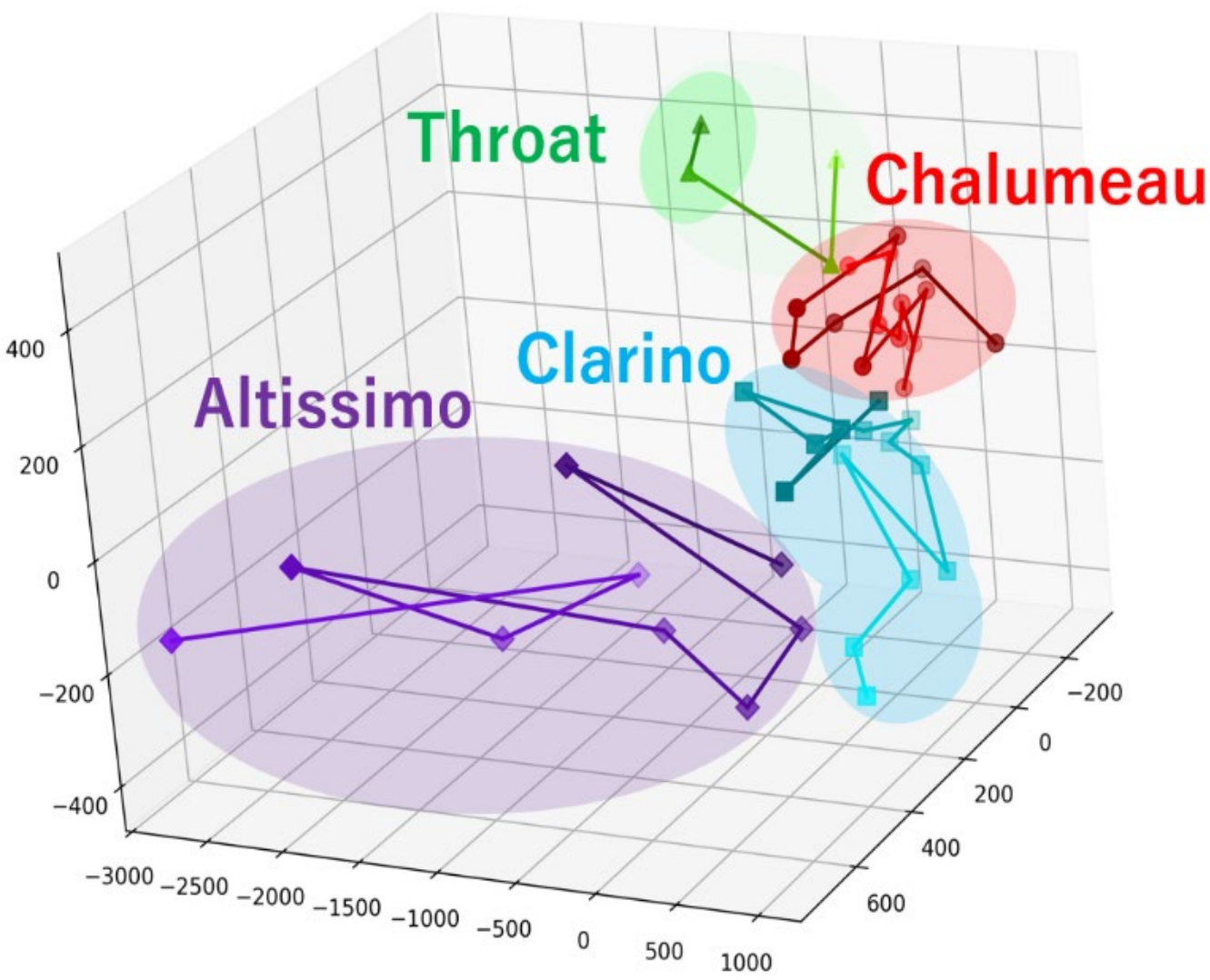

Fig. 7: An example of Topological Mapping based on the Wasserstein distance of the full-range spectrum of the B♭ clarinet.
It is possible to separate clusters corresponding to distinctive timbres.

It can be objectively confirmed that the spectrum in the vicinity of the "Throat G" note mentioned earlier exhibits a distribution that is distinctly separate from other timbre groups, from the low to the high frequencies. With regard to such notes that tend to stand out, it is well known amongst professionals within the woodwind section that there are techniques—such as 'alternative fingerings'—to cover this particular timbre. For example, in the bassoon section of the Vienna Philharmonic Orchestra, an 'alternative fingering' is traditionally used that differs from the standard fingering and is less likely to clash with the clarinet's "Throat G".

Let us consider a specific example from ensemble playing. Below is an excerpt from the first movement of Beethoven's Symphony No. 8. Here, whilst alternative fingerings do exist, the clarinet often adopts the unprotected "Throat G" fingering due to factors such as the musical phrase's dynamics; consequently, other instruments must compensate for this in order to achieve a high standard of ensemble playing.

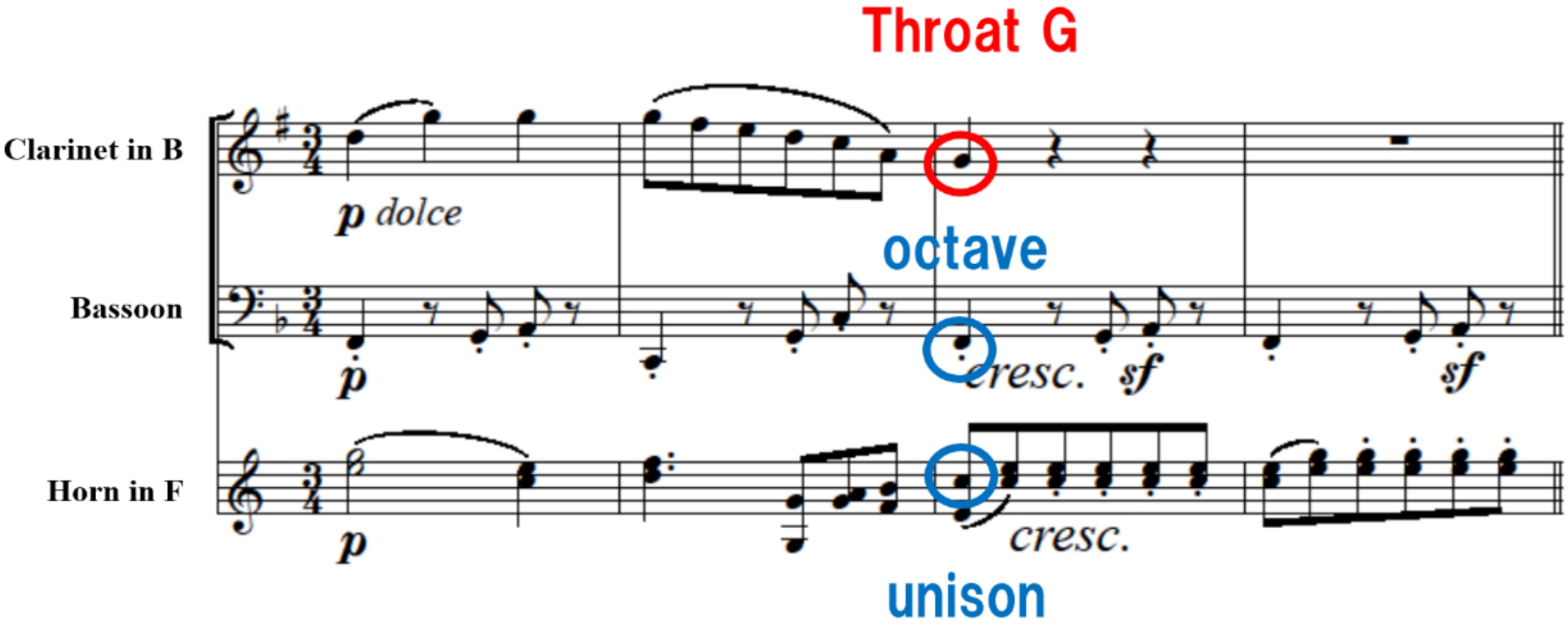

Fig. 7: An example of a difficult passage for the clarinet's 'Throat G' in an ensemble performance —Beethoven, Symphony No. 8

In the following, we shall compare the spectral characteristics of the 'alternate fingering' for the "Throat G" on the B♭ clarinet (which, as a transposing instrument, actually sounds as "F") with the spectral characteristics of the same F note (in unison) played on other woodwind instruments, using both standard and alternate fingerings.

In the case of the flute, there are few alternative fingerings for the lowest F; whilst the difference cannot be discerned simply by listening to the notes in isolation, a calculation of the Wasserstein distance suggests that the combination of the clarinet's "Throat G" and the flute's standard fingering may result in the smallest difference in spectral shape.

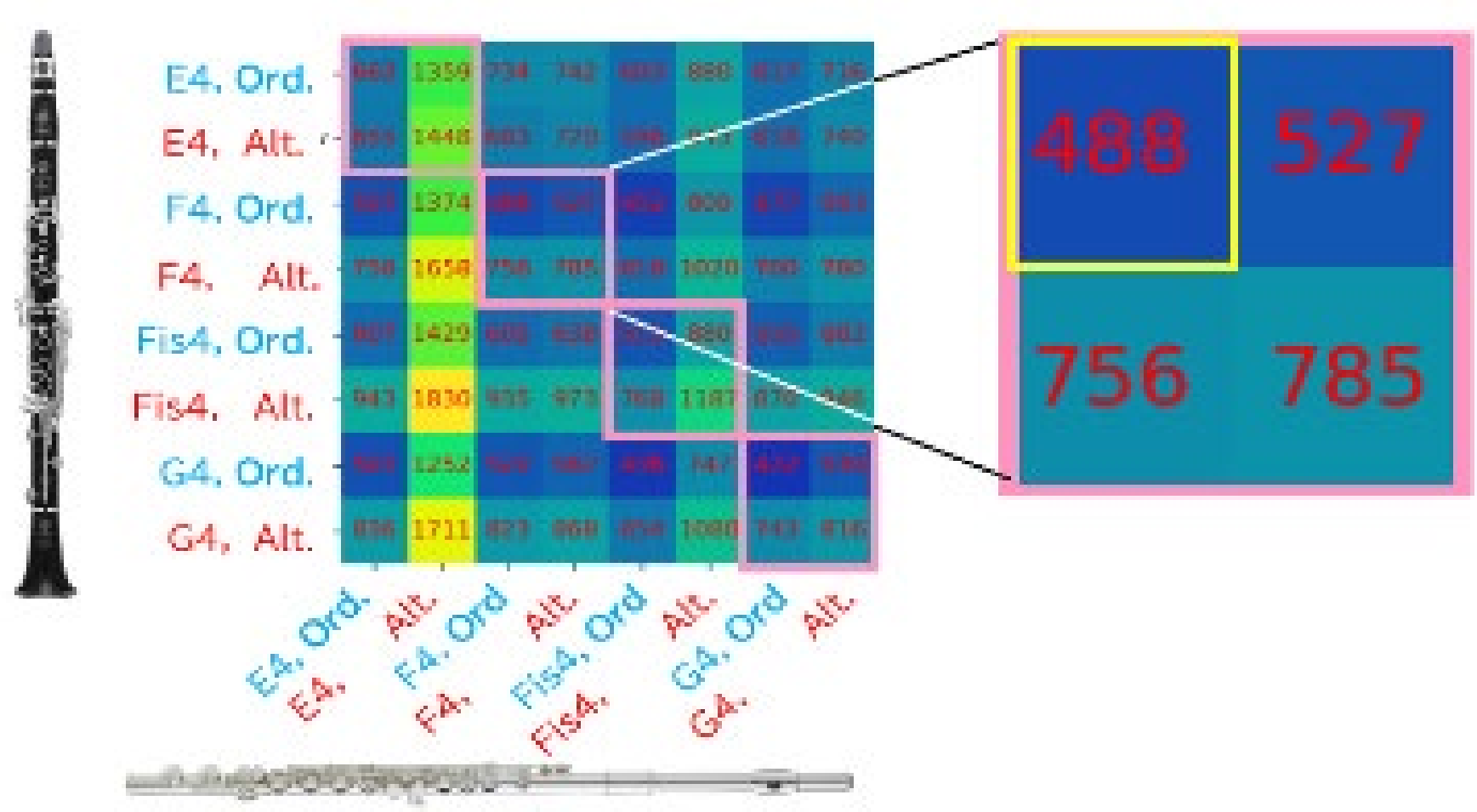

Fig. 8: Example of the compatibility between the clarinet's 'Throat G' and flute's unison

The situation differs for the oboe and bassoon compared to the flute due to the significantly different key systems. Compared to flute, double-reed instruments possess far more keys, and a wide variety of fingering possibilities can be envisaged, including those that are not normally used at all.

In this paper, in collaboration with the oboist Mr. Seiji TOMINAGA, of the Graduate School of Tokyo University of Fine Arts and Music, we devised a total of 13 new ‘F’ fingerings. We evaluated the Wasserstein distance between the spectra of all of these and the clarinet’s “Throat G”, thereby establishing a new fingering method with a notably high degree of compatibility.

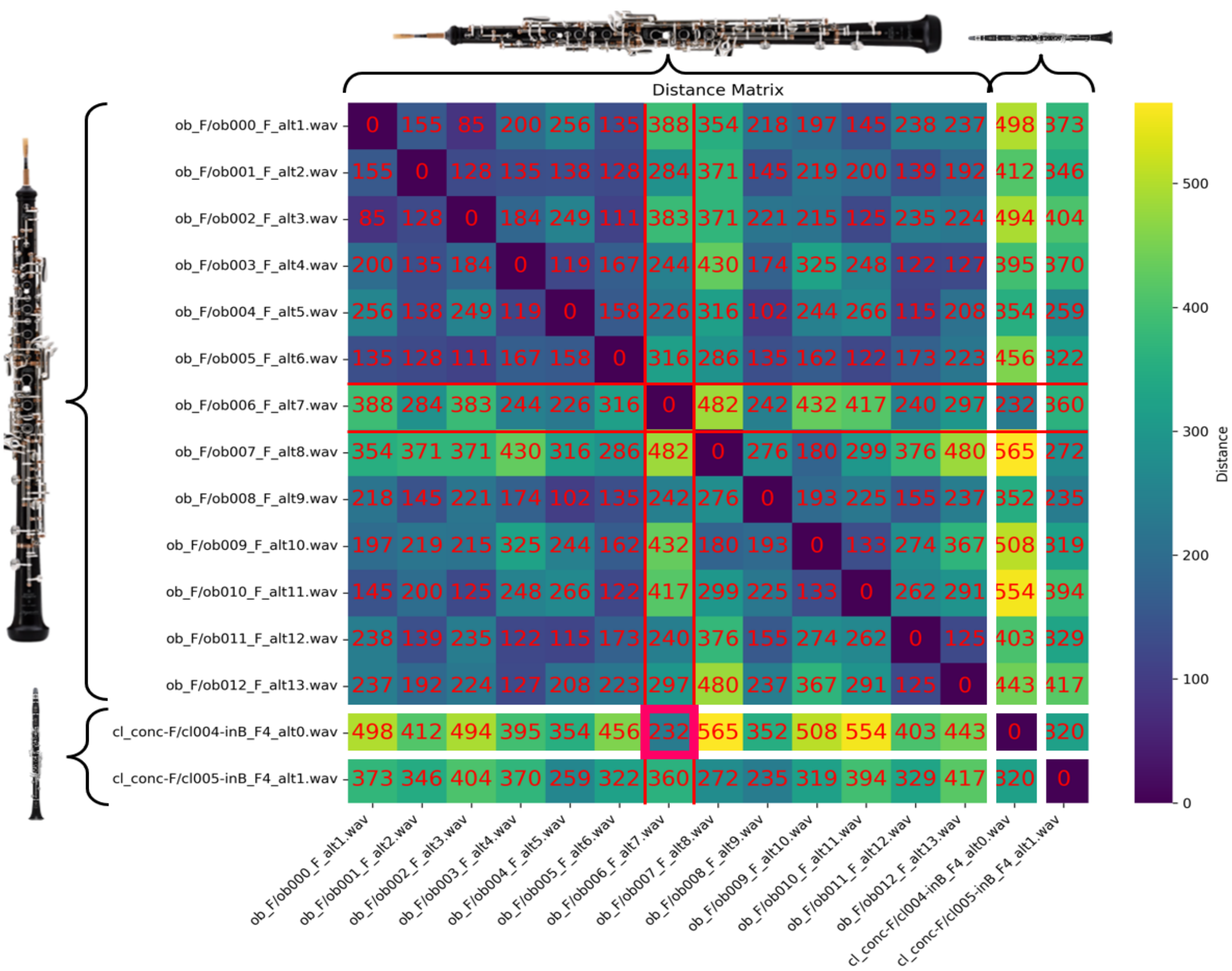

Fig. 9: The ‘Throat’ Sound of the Clarinet and the Unison Affinity of the Oboe

By taking the Wasserstein distance matrix as a starting point, we have been able to establish a framework for devising new fingering methods that contribute to improving ensemble quality, and this holds great promise for future developments.

With regard to the bassoon, with the cooperation of the fagottist Atsunori YUKIMASA of the Tokyo University of the Fine Arts and Music, we compared the spectra of alternative fingerings used by Stefan TURNOVSKY, Professor of Bassoon at the University of VIENNA—whom Mr. YUKIMASA studies under—with those of standard fingerings.

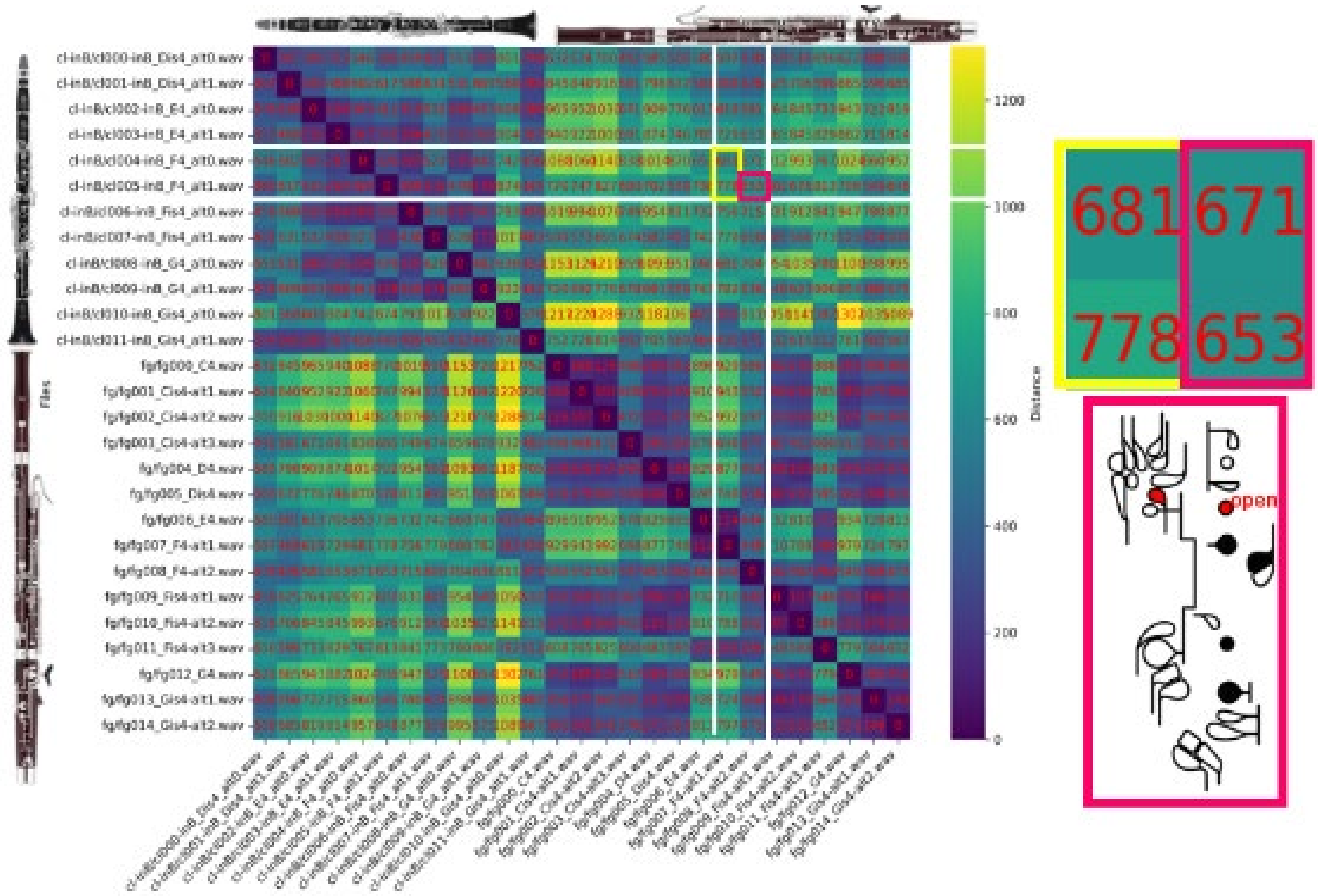

Fig. 10: The Throat G of B♭ Clarinet and the Unison affinity of Bassoon

It was confirmed that, in Professor TURNOVSKY's fingering, the Wasserstein distance was small regardless of whether the fingering for throat G of the clarinet was used or an alternative fingering was employed.

Professor TURNOVSKY joined The Vienna State Opera Orchestra in 1978 at the age of 19 and served as principal bassoonist at Vienna Philharmonic Orchestra for over 40 years; according to Mr. YUKIMASA, he is also well versed in various techniques for covering the weaknesses of other instruments in the woodwind section. The fingering described above is generally considered 'old fingering'—a type of so-called 'cross-fingering'—where a hole is opened partway along the tube.

Consequently, it is not always easy to achieve a stable tone. However, this allows the player to work behind the scenes to compensate for the weaknesses of other instruments in the section. It is of great value that such tips for optimal classical ensemble playing are not merely passed down orally amongst those who practice this tradition, but that their effectiveness is also confirmed through objective measurement. We should actively utilize this technique in the future to promote the wider adoption of such methods.

## 4. "Wasserstein deviation" in Klangfarbenakkord and the Progression of Klangfarbenharmonie

Thus far, we have considered the affinity or segregation of timbres based on the Wasserstein distance between two gamuts. In what follows, we shall consider quantification in terms of three voices (although not covered in this paper, similar methods can be applied to quartet ensembles and woodwind quintets; a separate paper dealing with more practical aspects of polyphony is currently in preparation).

In traditional harmony, a chord is represented by the concepts of "tension" and "resolution" (Cadence). However, in actual performance, Chords inevitably carry a specific timbre. This aspect has long been overlooked since classical antiquity; one of the clearest references to it in a musical context was Schoenberg's question, quoted at the beginning of this paper.

When considering the realities of ensemble playing, it is important to recognize that, in the formation of a harmonic texture, timbres that diverge drastically from those of the surrounding sounds will not be perceived as a unified "chord"; as demonstrated in the previous section, this requires performers to respond to one another in a flexible and adaptive manner. To quantitatively evaluate this 'unification of sound' amongst the constituent notes of a harmony, let us examine an indicator based on the Wasserstein distance.

Suppose the Wasserstein distances between three spectra constituting the harmony are given. Since the Wasserstein distance satisfies the triangle inequality, in most cases we can initially, quite naively, define a "Wasserstein Triangle" consisting of three-line segments that preserve the Wasserstein distance.

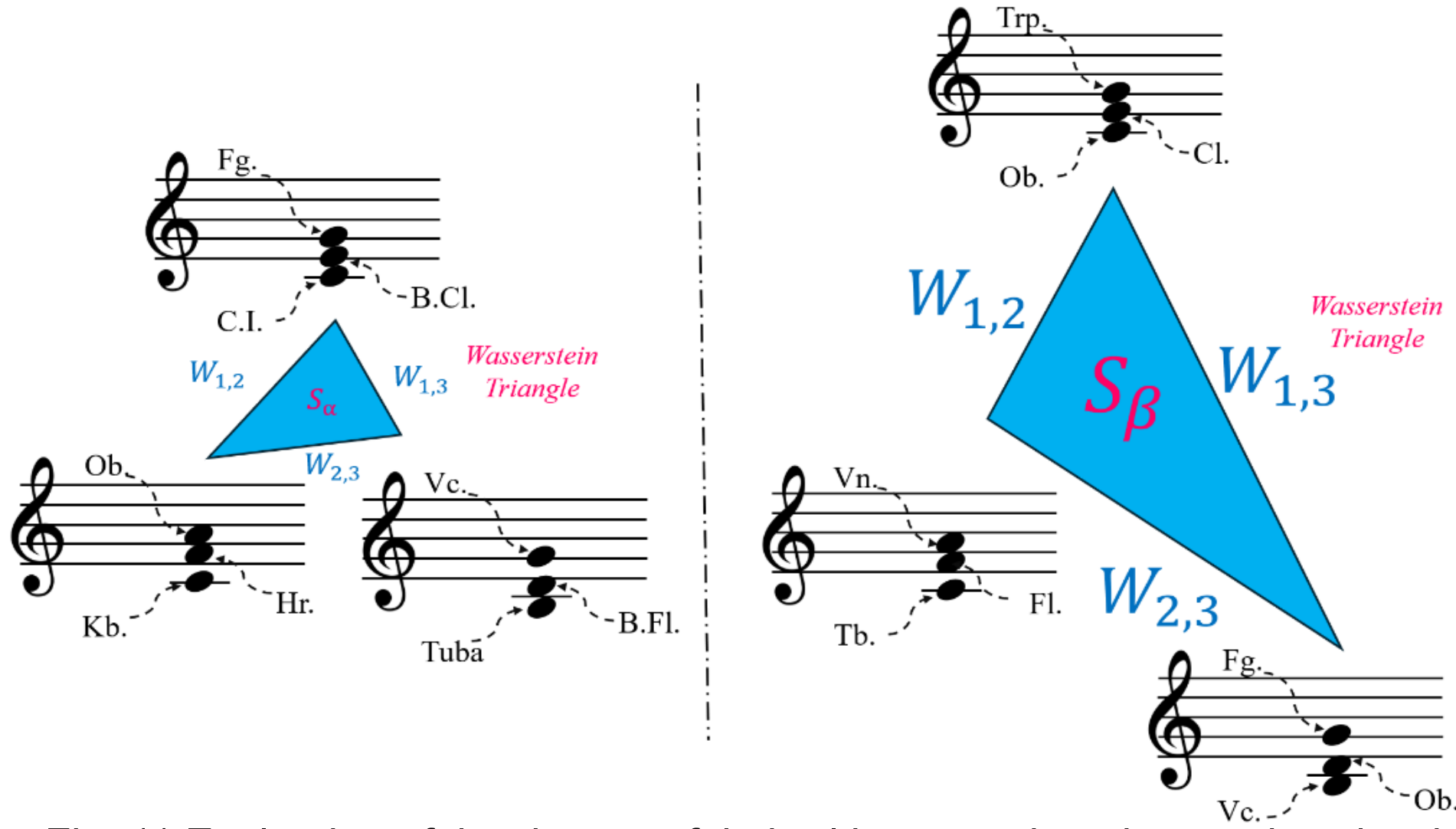

Fig. 11 Evaluation of the degree of timbral harmony in a three-voice chord using the 'Wasserstein triangle'. Schematic view

Based on the size of the area of this triangle, we can quantitatively assess the dissociation or affinity in the timbres of a “Klangfarbenakkord”. As the Wasserstein distance has the dimension [Hz], this area has the dimension [Hz²] or $\left[{}^{1}/_{\mathrm{sec}^2}\right]$.

Let us calculate the area of this triangle. When the Wasserstein distances $a$, $b$ and $c$ of the three sides forming the triangle are given, its area is calculated by

$$S = \sqrt{s\,(s-a)\,(s-b)\,(s-c)}, \quad s = \frac{a+b+c}{2}$$

(Heron’s formula). ・・・②

Fig. 12 shows the derived area of the Wasserstein Triangle for the instrument configurations indicated within the figure respectively. Taking the example on the far left, when close-position C chord is played with C on a Cor Anglais, E on a bass clarinet and G on a bassoon, the Wasserstein distances for the triad are calculated from the spectra of each instrument at their respective pitches, and the area is determined using Equation ②. This allows us to evaluate ‘Klangfarbenakkord’ indexed by various values, as shown in Fig. 12.

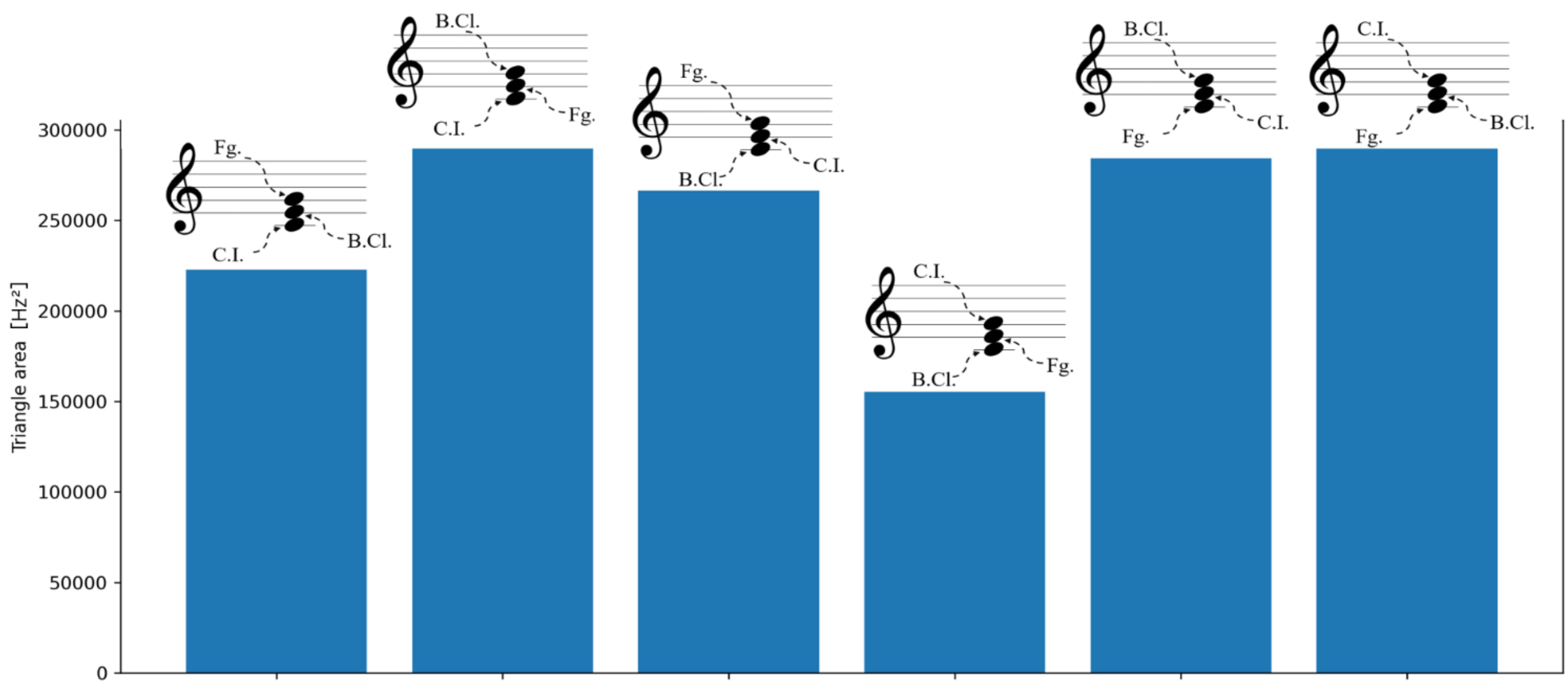

Fig. 12: An example of affinity evaluation based on the area of Wasserstein triangles

If we attempt to increase the number of voices in the ensemble and extend it to n parts ($n \geq 4$), it is generally impossible—as demonstrated above—to define a consistent $(n-1)$-dimensional pyramid and calculate the $(n-1)$-dimensional volume. In general, it is also impossible to construct a tetrahedron in three-dimensional space with four edges of arbitrary lengths, nor to construct a hyperprism of four or more dimensions with five or more edges of arbitrary lengths [8].

Let the three Wasserstein distances be $a$, $b$ and $c$ and the second moment $M_2$ (Wasserstein Timbre Variance) of the timbres, Wasserstein Timbre Deviation $\sigma = \sqrt{M_2}$ is defined as below.

$$M_2 = \frac{a^2 + b^2 + c^2}{3}, \quad \sigma = \sqrt{M_2} = \sqrt{\frac{a^2 + b^2 + c^2}{3}}. \qquad \cdots ③$$

Using equation ③ to calculate $\sigma = \sqrt{M_2}$ for the same set of chords as in Fig. 12 yields result that differ slightly from those in Fig. 12. Although this differs from area, it is possible to evaluate the timbral dispersion of triads. The dimension of $M_2$ is [Hz²], whilst that of σ is [Hz]; literally, these quantify the idiosyncrasy of spectral components.

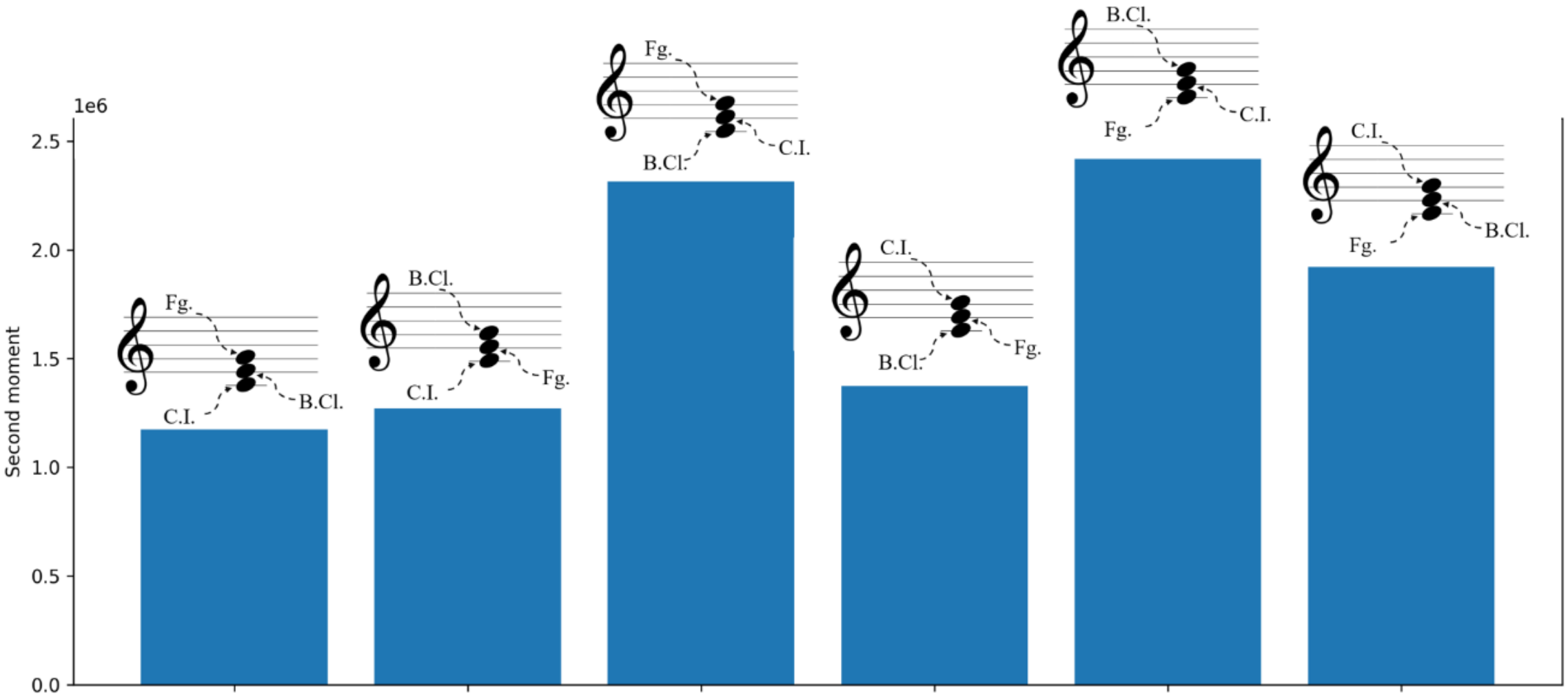

Fig. 13: An example of evaluating the timbral affinity of chords using Wasserstein deviation

When we re-examine these from the perspective of music, we notice an interesting fact. The ‘chords’ shown above all share the same function in the sense described above. Whilst the chord arrangement is the basic ‘close’ form, they can be arranged in either a ‘close’ or ‘open’ configuration as the first and second development forms. Furthermore, the ‘timbral dispersion’ should differ for each of these.

From this, it becomes possible to organize—or SERIALIZE —the instrumental Klangfarbenakkord arrangements, or ‘chord voicings’, which composers have traditionally orchestrated intuitively, according to an index of the degree of affinity or dissociation between the individual parts. If we consider ‘Klangfarbenakkord’ to be ‘affinitative’ in terms of Klangfarbenharmonie and ‘dissociate’ to be ‘independent as voices’, the following interesting structuring becomes possible.

Even if, in the traditional harmonic sense, the structure of tension and release is defined by cadences, it is possible to construct a ‘progression’ that, in terms of timbre, follows a path of unceasing ‘tension’; conversely, a progression that follows even a path of unceasing ‘realization’ is also possible.

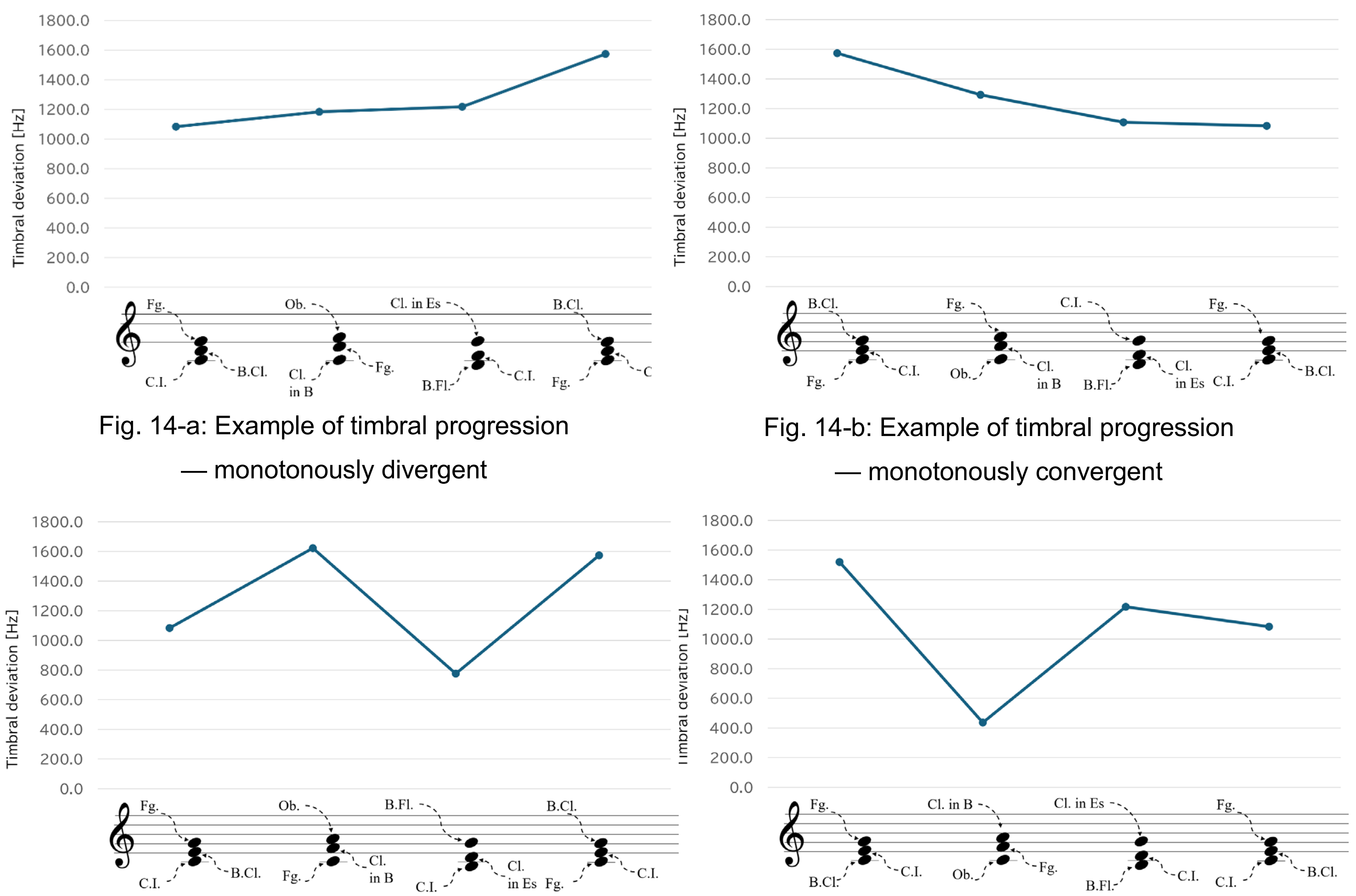

Fig. 14-a: Example of timbral progression — monotonously divergent

Fig. 14-b: Example of timbral progression — monotonously convergent

Fig. 14-c: Example of a timbre progression — parallel to the cadence

Fig. 14-d: Example of a timbre progression — contrary to the cadence

By following this procedure, it becomes possible to evaluate a ‘quantified timbral harmonic progression’ (quantifizierter klangfarbenharmonischer Verlauf) based on the quantification of individual ‘timbral chords’ (Klangfarbenakkord), thereby constituting a model within the field of ‘timbral harmony’ (Klangfarbenharmonielehre) based on the original idea of Arnold Schönberg.

As we are using the Wasserstein distance here, this system is mathematically “commutative”. If we were to conceive of quantification in terms of calculating conditional probabilities based on preceding chords, it would be possible to construct a different “Geometry of Harmony” — such as ‘sequential timbral harmony’ (sequenzielle Klangfarbenharmonielehre) — which involves irreversibility=incommutativity, utilizing measures such as KL divergence.

## 5. From “die Klangfarbe” to “die Sprechstimme”

It is reasonable to consider that the scientific approach to human speech and instrumental timbre originated in the 1850s with Hermann von Helmholtz’s work on physiological acoustics and vowel synthesis [9]. Whilst the spectral analysis of instrumental timbres can be traced back to the work of Lord Rayleigh [10] and Chandrasekhar Raman’s research into musical instrument timbres [11], the analysis of speech did not begin until the Second World War, with the development of the vocoder by Dudley and his colleagues at AT&T Bell Laboratories [12].

As for speech synthesis, apart from the work of Karl Willy Wagner (1936) [13]—who replaced Helmholtz’s vowel tuning forks with an electrical oscillator—a systematic approach did not emerge until Gunnar Fandt’s “source-filter theory” [14], which produced vowel synthesis (OVE1, 1953) and speech synthesis including consonants (OVE2, 1962).

These achievements were significantly influenced by military technology developed during the Second World War, such as the encryption of the hotline linking the White House and London; the PCM (Pulse Code Modulation) established at that time forms the foundation of all modern digital audio processing.

Around the same time as the announcement of the Mark I analogue synthesizer by Harry Olson and Herbert Beller at the RCA Princeton Laboratory [15], it became possible to synthesize musical instrument timbres; however, early analogue synthesized sounds possessed a distinctive timbre (reminiscent of electronic organs, for example) and were a far cry from the sounds of real musical instruments.

The synthesis of timbres resembling those of physical instruments became more active in the 1970s, following the development of FM synthesis by John Chowning and others towards the end of the analogue era [17], and eventually gained momentum in tandem with digitalization through the advent of sampling synthesizers [18] and physical modelling synthesis [19].

In tandem with these developments, “timbre mapping” using MDS (Multidimensional Scaling) emerged, which treats the sounds of real instruments as cognitive entities and maps them according to their degree of similarity or dissimilarity [20], C. Krumhansl and others presented research on timbre mapping [21], driven by the simplification of digital audio processing and the rise of personal computers in the 1980s; however, as the quality of digital synthesizers stabilized in the 1990s, research demand within the fields of science and engineering began to wane [22].

With the rise of machine learning-based sound generation in the 2010s, research into the cognitive features of timbre by C. Agon and Ph. Esling, along with its geometrization, emerged [23], became more sophisticated, and progressed towards the creation of “Generative Timbre Space”—a morphing technique that uses “variational autoencoders” to seamlessly connect existing instrumental timbres.

Esling et al.’s approach is highly practical from a musician’s perspective. By regularizing the latent space using psychoacoustic timbre similarity, it continuously interpolates and generates timbres across different instruments; this approach can be seen as a culmination of the system pioneered in the 1950s by Pierre Boulez, who conceived the “serialization of timbre” [24] and realized it in instrumental works such as “Marteau sans maître” (1955), now brought to electronic perfection.

Boulez, who had a background in mathematics, tended to classify various musical parameters according to “striated and smooth” (Strié et lisse), that is, in terms of the continuum and the discrete. The approach taken in works such as Agon and Esling can be understood as treating existing instrumental timbres as ‘lattice points’, so to speak, and by smoothly interpolating between them, expanding the timbral manifold of musical expression into a single continuous entity.

On the other hand, our group is conducting research from the perspective of musicians and composers trained in cognitive psychology and physics, focusing on different areas of interest such as the development of new playing techniques, the improvement of ensemble quality, and the exploration of new compositional principles; we are thus adopting an approach based on distinct interests that, in a positive sense, complements the body of previous research.

In this study, we have not identified representative timbres of instruments corresponding to ‘lattice points’. Rather, by evaluating the minute characteristics of the information manifold of instrumental acoustics not in terms of the psychological properties of individual timbres, but in terms of the physical optimal transport distances between multiple timbre spectra, we aim to make the various quantities associated with ensemble playing computable, thereby taking a step forward towards resolving Schoenberg’s classic question.

According to Luigi NONO, total serialism in the 1950s was exploring the following three sets of Questions by A. SCHÖNBERG. [25].

1. How should sounds (sequences) be organized?
2. What is the difference between speaking and singing?
3. New developments arising from timbre and melody

In addition to these, attention was drawn to how to address new parameters in sound processing via electrical and electronic media—made possible by post-Second World War technological advances—such as the issue of spatiality. A glance at the later works of Nono, BOULEZ and STOCKHAUSEN — who were opinion leaders in the 1950s—reveals the three distinct approaches each took in responding to “SCHÖNBERG’s 3 questions”.

BOULEZ composed and performed Répons (1981/84), a work without vocals that integrated digital sound processing into the system; NONO, in Prometeo (1984) for instrumental music and voice accompanied by electronics, presented a ‘de-construction’ within a spatial theatre without a center;

STOCKHAUSEN, who had been conscious of spatiality from an early stage, dating back to his early "Gruppen" (1955), undertook a variety of experiments in "Licht" (1977–2003); in particular, in the electronic music section Oktophonie (1990–91) from the second act of 'Tuesday', he succeeded to a certain extent in achieving vertical movement, in addition to the two-dimensional movement of the electronic soundscape characteristic of BOULEZ and NONO [26].

Over the past thirty-odd years, our research group has approached 'SCHÖNBERG's three Questions' independently, getting distinct results for each. However, from 2025 onwards, we realized that by utilizing the framework of Information Geometry proposed by Professor Shunichi AMARI, we could reframe these three themes as a single system of problems.

If the spectra of speech and music are regarded as probability density functions, it is possible to calculate 'spectral series' relating to 'timbre', 'speech' and even their 'coupling' by performing appropriate operations on the relevant quantities. This makes it possible to construct a general-purpose information geometry for music.

Rather than treating SCHÖNBERG's three problems independently, our approach explores a triadic information-geometric structure—comprising 'Series of sounds', 'Series of timbres' and 'spoken language as dynamical variation in timbre'—by combining them in a mathematical product. To put it succinctly, one might say: 'From "die Klangfarbe" to " die Sprechstimme".

## 6. Discussion and Concluding Remarks … "Ehrenfest's Theorem" in Klangfarbenharmonielehre

Let us point out an important fact to conclude; the <Ehrenfest Theorem> also holds true in "Klangfarbenharmonielehre". In 1927, the physicist Paul Ehrenfest proved that the results (expectation values) of Quantum Mechanical calculations dealing with microscopic systems will always be in accordance with those of Newtonian mechanics when taken to the macroscopic limit.

Similarly, based on the examples in this paper, the limit where the various Wasserstein metrics asymptotically approach zero—that is, the limit where the difference in timbres degenerates—leaves classical music theory, such as harmony, intact.

$$\lim_{\text{Wasserstein index} \to 0} F_{\text{Klangfarbenharmonie}} \rightarrow F_{\text{Klassische Harmonie}} \cdots ④$$

This fact guarantees that this approach is not limited to a specific musical genre; rather, it opens up possibilities for broad application—from commercial pop music and traditional ethnic music to cultural rituals and various contexts related to speech and sound. This musical utility, regardless of genre, is precisely what Max Mathews emphasized from the very early days of computer applications in music.

# Notes and references

[5] Let $\mu_f$ and $\mu_g$ be the centers of mass of two probability density functions $f$ and $g$ (whose spectra are normalized to 1 under the total measure). With regard to the one-dimensional Wasserstein distance $W_1(f,g)$ and the center-of-mass shift $|\mu_f - \mu_g|$, the following holds constantly:

$$W_1(f,g) \geq |\mu_f - \mu_g|$$

Using the cumulative distribution functions $F(x) = \int_{-\infty}^{x} f(t)\,\mathrm{d}t$ and $G(x) = \int_{-\infty}^{x} g(t)\,\mathrm{d}t$, the one-dimensional Wasserstein-1 distance can be expressed as

$$W_1(f,g) = \int_{-\infty}^{\infty} |F(x) - G(x)|\,\mathrm{d}x\ ,$$

then, by the integral formula for absolute values and the triangle inequality for any measurable function $h(x)$,

$$W_1(f,g) \geq \left| \int_{-\infty}^{\infty} \big(F(x) - G(x)\big)\,\mathrm{d}x \right| .$$

From the properties of probability density functions, the following holds between the center of mass (mean) and the integral of the cumulative distribution function:

$$\int_{-\infty}^{\infty} \big(F(x) - G(x)\big)\,\mathrm{d}x\ = \mu_f - \mu_g$$

holds true; therefore, substituting this immediately yields

$$W_1(f,g) \geq \left|\mu_f - \mu_g\right| .$$

As stated in the text, we can interpret the difference between the two, $\Delta W$, as the ‘equivalent change in timbre’.

[6] The reasons why such an approach has been rarely employed in the analysis of instrumental sounds and speech include the fact that discussions on optimal transport were only introduced into audio processing from the 2010s onwards, and that their application has been limited to engineering fields such as machine learning, signal processing and source separation; furthermore, as metrics for representing differences in timbre, KL divergence, Euclidean distance, or distances in MFCC (Mel-frequency cepstral coefficients) space incorporating the perceptual characteristics of human hearing.

In this study, with regard to changes in physical sound waves prior to the application of a perceptual filter, we geometrically decompose these into ‘centroid shifts (changes in pitch/brightness)’ and ‘changes in higher-order structure (such as the transition between chest voice and head voice)’, and by determining and defining them as scalar quantities with physical meaning—namely, ‘the amount of change in timbre structure [Hz]’—we have obtained musically meaningful indicators.

We have established this new framework because it is difficult to obtain results that are effective for musical purposes from the traditional perspectives of acoustic engineering and similar fields.

[7] We are conducting an analysis in which we select three eigenvectors in descending order of their positive eigenvalues and embed the data into a space spanned by these eigenvectors; however, as the parameter space exhibits curvature, we are currently preparing to investigate the Riemannian manifold.

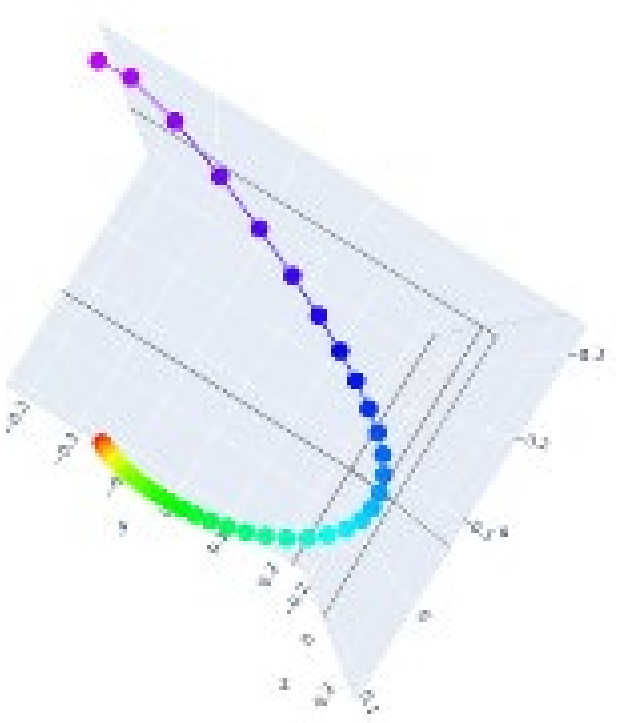

Fig. 15-a Eigenvector mapping of Gaussian noise.

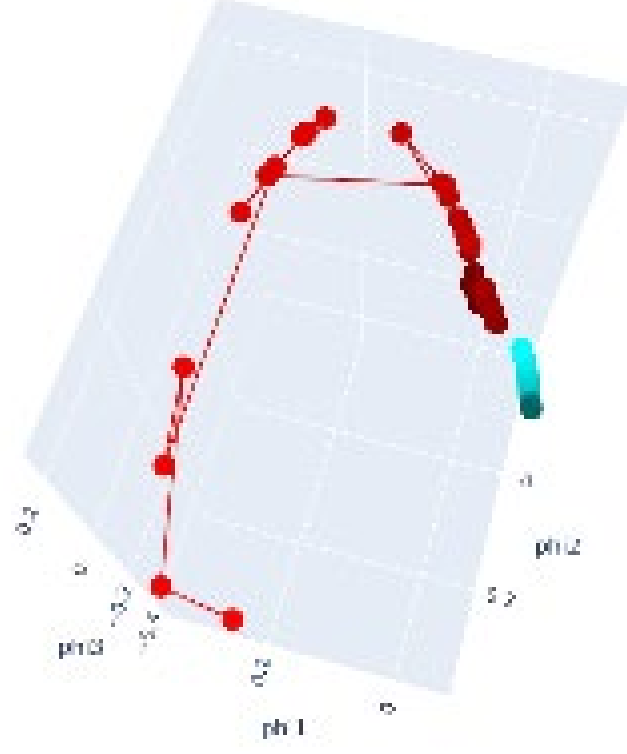

Fig. 15-b A similar example of a flute accompanied by Gaussians.

The example on the left shows a case where we prepared Gaussian noise with the same pitch as the instrument under consideration and expanded it in the eigenvalue space using only that noise, thereby revealing the curvature of the space.

However, in the case where the flute’s scale has been co-diagonalized (right), the sequence of Gaussian scales is reduced to a series of blue dots, demonstrating that the spatial distortion inherent in the flute’s timbre is far greater. I shall elaborate on the details in a subsequent article.

[8] Given $N+1$ line segments in an $N$-dimensional space, the necessary and sufficient condition for them to form a simplex (an $N$-dimensional cone) is the existence of a combination of angles (internal angles) formed by adjacent line segments at every vertex; in practice, this can be assessed using the Gram matrix. Since the Wasserstein distance obtained from any set of spectra does not necessarily satisfy the conditions defined by the Gram matrix, it is not possible to construct an evaluation method using area or volume for dimensions of four or higher.